\documentclass[pra,twocolumn]{revtex4}

\usepackage{graphicx}
\usepackage[usenames,dvipsnames]{color}

\begin{document}

\title{Ab initio properties of Li-Group-II molecules for ultracold matter studies}

\vspace*{0.5cm}
\author{Svetlana Kotochigova}
\email[Corresponding author: ]{skotoch@temple.edu}
\affiliation{Department of Physics, Temple University, Philadelphia, PA 19122, USA}
\author{Jacek K{\l}os}
\affiliation{Department of Chemistry and Biochemistry, University of Maryland, College Park, MD 20742, USA}
\author{Alexander Petrov}
\altaffiliation{Alternative address: St. Petersburg Nuclear Physics Institute, Gatchina, 188300;
Department of Physics, St.Petersburg State University, 198904, Russia}
\author{Maria Linnik}
\affiliation{Department of Physics, Temple University, Philadelphia, PA 19122, USA}
\author{Paul S. Julienne}
\affiliation{Joint Quantum Institute, NIST and University of Maryland, Gaithersburg, MD 20899-8423, USA}

\begin{abstract}
We perform a systematic investigation of the electronic properties
of the $^2\Sigma^+$ ground state of  Li-alkaline-earth dimers. These
molecules are proposed as possible candidates for quantum simulation of
lattice-spin models. We apply powerful quantum chemistry coupled-cluster
method and large basis sets to calculate potential energies and permanent
dipole moments for the LiBe, LiMg, LiCa, LiSr, and LiYb molecules. Agreement of calculated 
molecular constants 
with  existing experimental data is better than or equal to 8\%.
Our results reveal a surprising irregularity in
the dissociation energy and bond length with an increase in the reduced
mass of the molecule. At the same time the permanent dipole moment at the
equilibrium separation has the smallest value between 0.01 a.u.  and 0.1
a.u. for the heaviest (LiSr and LiYb) molecules and increases to 1.4 a.u.
for the lightest (LiBe), where 1 a.u. is one atomic unit of dipole moment.
We consider our study of the $^2\Sigma^+$ molecules
a first step towards a comprehensive analysis of their interactions in
an optical trap.  
\end{abstract}

\maketitle

\section{Introduction}

There is growing interest in molecules formed by
one alkali-metal atom and one alkaline-earth or rare-earth atom
\cite{Zoller2006,Carr2009,Krems2010,sorensen,Okano2010,Hutson,Dulieu2010,Zhang,Gopakumar2010,Gupta2011,Pashov}.
In contrast to group I or group II di-atomic molecules, these
molecules have an unpaired electron that allows for manipulation of the
X$^2\Sigma^+$ ground-state molecule with both electric and magnetic
external fields. Moreover, it has been suggested that $^2\Sigma^+$
molecules are good candidates in which to explore controlled chemical
reactions at ultracold temperatures \cite{Krems2006}. The unpaired spin in the ground state of these
molecules should provide a new handle for the control of the reaction
dynamics based on the interplay between intramolecular spin-rotation
couplings. In addition, when placed in an optical lattice,
these polar molecules can interact with each other via both electric
dipole-dipole and magnetic spin-spin forces. It is therefore possible
to engineer unusual forms of interactions for quantum simulation of
lattice-spin models and topological quantum computing \cite{Zoller2006},
controlled preparation of many-body entangled states \cite{Krems2010}, and
high-precision measurements of fundamental constants \cite{Hutson2002}.

Another interest in Li-group II  molecules stems from prospects to
achieve optical Feshbach tuning of scattering properties in ultracold
gases of the individual atoms without substantial loss of atoms
from a trap. This photoassociative tuning, pioneered for homonuclear
molecules \cite{Fedichev1996,Bohn1996,Fatemi2000,Theis2004,Ciurylo2005,Thalhammer2005,JonesReview2006,Enomoto2008},
becomes possible due to the existence of long-lived excited molecular
states near narrow intercombination lines of the alkaline-earth or
rare-earth atoms.  In turn, this might enable efficient ways to form
gases of polar molecules without substantial loss by photoassociation. 
In essence, a two-photon
optical Feshbach resonance can be used to couple two colliding atoms to
a vibrational level of the molecular ground state. The suppressed effect
of excited-state spontaneous decay makes efficient coherent molecular
formation possible. Knowledge of the electronic and ro-vibrational
properties of  the molecules will help to find optimal pathways.
Characteristics such as the permanent dipole moments of the ground state
will determine the anisotropic interactions between these molecules in
an optical trap.

Some key theoretical predictions of the electronic structure and
dipole moment of the ground and low-lying excited states of  LiYb
\cite{Zhang,Gopakumar2010} and  for the ground state of LiSr
\cite{Dulieu2010}  have recently been obtained. Other work has
studied RbYb \cite{sorensen} and RbSr \cite{Hutson}.  However, the
physical origin of the chemical bond of these mixed species  is not
yet fully understood. Continued theoretical advances are necessary to
unravel the properties of these molecular systems.  To the best of our
knowledge, only the LiMg and LiCa ground states were experimentally
investigated \cite{Pichler,Berry,Pashov}, which allows us to compare
our predictions with experimentally obtained molecular parameters
\cite{Berry,Pashov}.  There are several experimental groups working to
achieve the formation of the LiYb molecules at  ultracold temperatures
\cite{Okano2010,Gupta2011}. An understanding and quantitative description
of these molecules might help to define conditions  under which these
ultracold molecular systems can be created.

Our theoretical study is devoted to a systematic investigation of the
X$^2\Sigma^+$ ground state properties of the polar Li-alkaline-earth
dimers such as LiBe, LiMg, LiCa, LiSr, and the rare-earth dimer LiYb.
These are interesting and challenging systems as the valence electrons of
the atoms form a relatively weak covalent bond  \cite{Bauschlicher}. The
alkaline-earth atom has a closed outer electron shell and a Hartree-Fock
interaction energy results in a purely repulsive potential curve. We use
the unrestricted (U) or partially spin-restricted (R) coupled cluster
(CC) method with either single and  double excitations augmented with
pertubatively calculated triples (CCSD(T)) or with single, double,
and triple excitations (CCSDT) to treat the correlations of core
and valence electrons. All approaches are used in conjunction with
correlation-consistent basis sets. The dipole moment is calculated using
a finite field approach with a four-point numerical differentiation
formula. The calculations are performed with CFOUR \cite{CFOUR} and MOLPRO
\cite{MOLPRO} suites of programs.

Presumably, the CCSDT method is more accurate in the description of the
potential energy surface than the CCSD(T) method, where triple excitations
are treated perturbatively.  The results of the unrestricted (U) method
are more accurate than that of a restricted (R) calculation.
The CC method is based on a single-reference configuration and is not
suitable for molecules where multiple-configurations are relevant. A
practical indicator of how well the CC method performs is based on T1
diagnostics \cite{Lee1989}. Our calculation shows that T1 is small and
varies from 0.015 to 0.035. This indicates that a single-reference method
is valid for these dimers.  In fact, a small T1 for all interatomic
separations implies that the long-range interaction potentials as well
as the dissociation energies can be accurately described.

Our results include the ground-state potential curve, bond distance,
harmonic frequency, dissociation energy, and permanent dipole moment
for  LiBe, LiMg, LiCa, LiSr, and LiYb. For completeness, we calculate
vibrational energies of the ground-state potentials. 
Due to limitations in computational resources  we were not able to obtain
results for all methods, basis sets, and Li-X diatoms.  The most accurate
potential energy curves and dipole moments are presented in the figures
and tables with the exception of Tab.~\ref{all_const}, which compares
molecular constants obtained with the various methods and basis sets.

All potentials are corrected for basis set superposition errors (BSSE)
using the counterpoise procedure of Boys and Bernardi \cite{Boys:70}.
The sign of the dipole moment depends on the position of the
origin.  We apply the following convention throughout the paper: The
alkaline-earth/rare-earth atom is positioned at the origin and the Li atom
moves along the positive $x$-axis.  A positive sign of the dipole moment
corresponds to the charge transfer from the Li atom towards the alkaline-earth/rare-earth
atom. 

\subsection{LiBe molecule}

There exist several theoretical calculations of the X$^2\Sigma_{1/2}^+$
electronic ground state of the LiBe molecule. Jones \cite{Jones} used a
density functional formalism based on a local density approximation to
predict molecular constants of the LiBe, LiMg, and LiCa molecules. Their
dissociation energy is  $D_e/(hc)$ = 3705.9 cm$^{-1}$, while
the equilibrium bond length $R_e$ is $4.9 a_0$ and the vibrational
constant $\omega_e/(hc)$ = 340 cm$^{-1}$. Such strong bonding of the LiBe
molecule was explained by substantial electron-charge transfer between
s and p orbitals.  More advanced MRCI calculations of the low-lying
doublet and quartet states of LiBe was performed by Fischer {\it et al.}
\cite{Fischer1991}. In this calculation the ground state is bound by
2497.5 cm$^{-1}$, $R_e = 4.95 a_0$, and $\omega_e/(hc)$ = 302 cm$^{-1}$. The
permanent  dipole moment is  $d_e$ = 0.95 a.u. at $R_e$.
Another MRCI calculation \cite{Bauschlicher} using all single and double
excitations  obtained $D_e/(hc)$ = 2336.35 cm$^{-1}$, $R_e = 4.927 a_0$,
and $\omega_e/(hc)$ = 300 cm$^{-1}$, which are close to those of Ref.~\cite{Fischer1991}.  
However, the dipole moment
$d_e = 1.33$ a.u. of Ref.~\cite{Bauschlicher} differs significantly
from that of Ref.~\cite{Fischer1991}. The effect of inner-shell
correlations on the molecular electronic structure was investigated.
Later an {\it ab~initio} configuration-interaction study was reported
by Marino and Ermler \cite{Marino1992}. Their value of the dissociation
energy $D_e/(hc)$=2014 cm$^{-1}$ is smaller than that in previous studies.
In summary, the previous theoretical calculations disagree on dissociation
energy and dipole moment values but have much better agreement on the
value of the bond length. Our goal is to test results of the previous
predictions by performing a coupled cluster calculation accounting for
high-order correlation effects.

\begin{figure}
\includegraphics[scale=0.4]{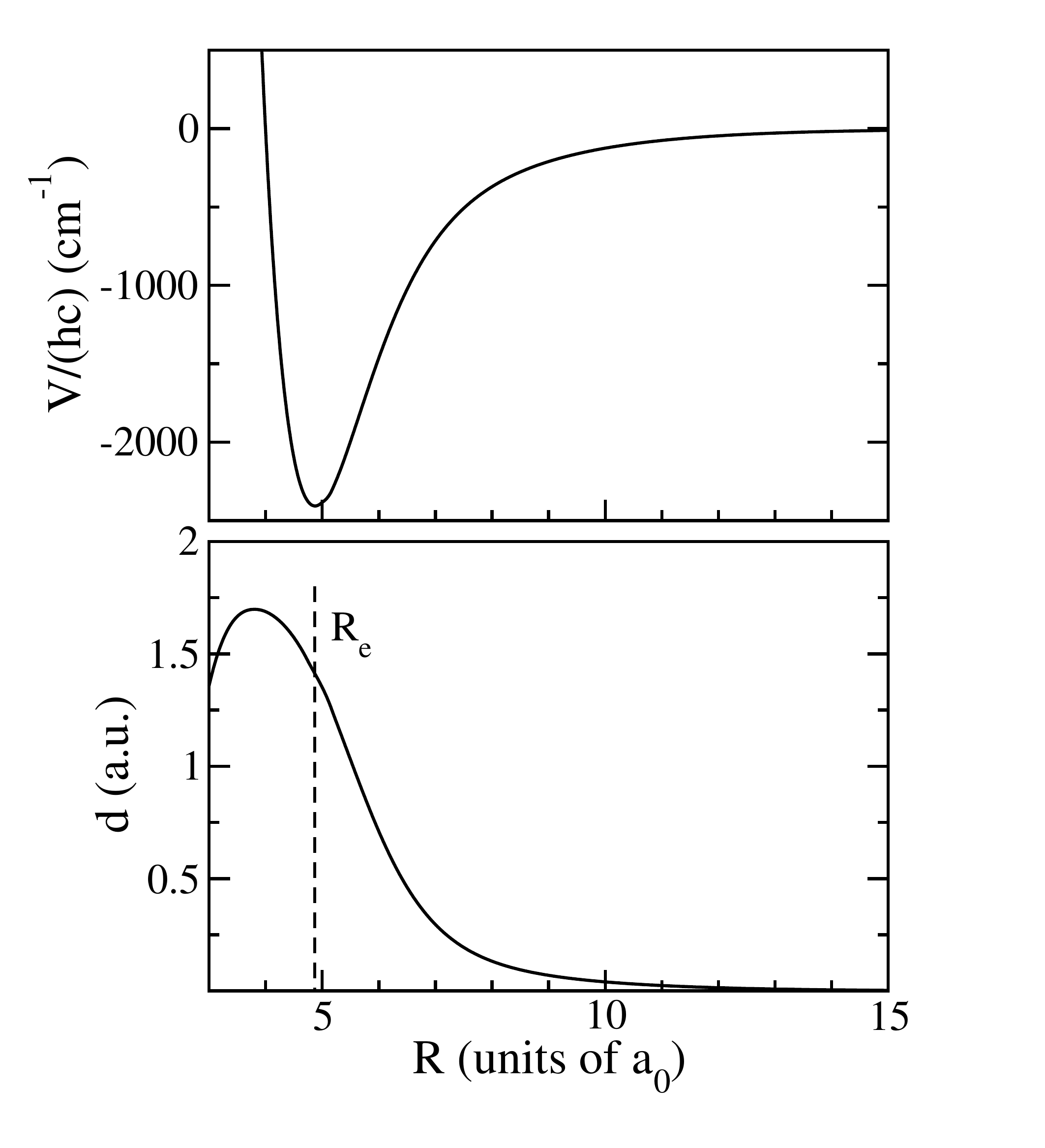}
\caption{Top panel: electronic $^2\Sigma^+$ ground-state potential of
LiBe; bottom panel: permanent dipole moment function of the ground state
of LiBe.  The vertical dashed line indicates the equilibrium separation of
the ground state potential, at which the dipole moment equals 1.41 a.u.
The vibrationally-averaged dipole moment of the $v=0$ level is equal to a
slightly smaller value of 1.372 a.u.  
 }
\label{li_be}
\end{figure}

In our calculation the UCCSD(T) method is used to obtain the ground
state potential and dipole moment of LiBe as a function of interatomic
separation $R$. These characteristics are calculated  with the large
basis set (aug-cc-pV5Z-DK) of Prascher {\it et al}~\cite{Prascher},
which includes scalar relativistic effects accounted for by the
Douglass-Kroll Hamiltonian. Atomic basis sets include (15s8p5d4f3g2h) primitive Gaussian
and [7s6p5d4f3g2h] constructed functions for Li and (15s9p5d4f3g2h)/[7s6p5d4f3g2h] 
for Be. 

Results of our calculation are shown in Figure~\ref{li_be}.  Comparing our
data with results available from other theories and experiment we conclude
that $R_e = 4.87 a_0$ is close to the experimental measurement of $R_e
= 4.894 a_0$ \cite{Schlachta1990}, while the well depth $D_e/(hc)$
= 2406.89 cm$^{-1}$ is in best agreement with the prediction of
\cite{Bauschlicher}. In Ref.~\cite{Schlachta1990} no value of $D_e$
is given.  The dipole moment value $d_e$ = 1.41 a.u. at the equilibrium
separation is even larger than that of \cite{Bauschlicher}.  The permanent
dipole moment averaged over the ground state vibrational wave function
is equal to 1.37 a.u. It is slightly lower than the value of the dipole
moment function at $R=R_e$.  The LiBe molecule has the largest dipole
moment of the molecular systems considered in this work.

Bound states of $^7$Li$^9$Be are calculated using a discrete variable
representation (DVR) \cite{Colbert}, and their binding energies are presented
in Table~\ref{libebound}.  The zero-point energy
is found to be $D_0/(hc)$ = 2254.29 cm$^{-1}$.

\begin{table}
\caption{All vibrational energies of the $^7$Li$^9$Be X$^2\Sigma^+$ ground state potential 
for the rotational state $N=0$.} 
\begin{ruledtabular}
\begin{tabular}{rcrc}
$\nu$&Energy/($hc$) (cm$^{-1}$)& $\nu$&Energy/($hc$) (cm$^{-1}$)\\
0   &  -2254.29 &9  &   -337.24\\
1   &  -1954.76 &10 &   -243.42\\
2   &  -1669.25 &11 &   -169.45\\
3   &  -1407.21 &12 &   -112.19\\
4   &  -1167.64 &13 &    -68.88\\
5   &   -951.37 &14 &    -37.48\\
6   &   -760.21 &15 &    -16.55\\
7   &   -594.38 &16 &     -4.86\\
8   &   -453.64 &17 &     -0.50\\
\end{tabular}
\end{ruledtabular}
\label{libebound}
\end{table} 

\subsection{LiMg molecule}

To the best of our knowledge, there exist two experimental
investigations of the LiMg ground state properties 
\cite{Pichler,Berry}. The study in Ref.~\cite{Berry} provides
molecular constants that we can use for comparison with our
calculation. This work was based on resonant photo-ionization
spectroscopy of LiMg in the gas phase and gives an accurate bond
energy $D_0/(hc)$ = 1330 cm$^{-1}$, equilibrium separation $R_e=5.9
a_0$,  and vibrational frequency $\omega_e/(hc)$ = 190 cm$^{-1}$.
A density functional study in Ref.~\cite{Jones}
of LiMg gives $D_e/(hc)=1371$ cm$^{-1}$, $R_e=6.04
a_0$, and $\omega_e/(hc)$ = 180 cm$^{-1}$.  A MRCI approach with large
contracted basis sets and multiple sets of polarization functions is
used in Ref.~\cite{Bauschlicher} to give predictions for the LiMg ground
state molecular constants $D_e/(hc)= 1611.28 $ cm$^{-1}$, $R_e= 5.88  a_0$,
and $\omega_e/(hc)$ = 183 cm$^{-1}$. The value for $D_e$ significantly differs
from both Ref.~\cite{Jones} and experiment.

\begin{figure}
\includegraphics[scale=0.4]{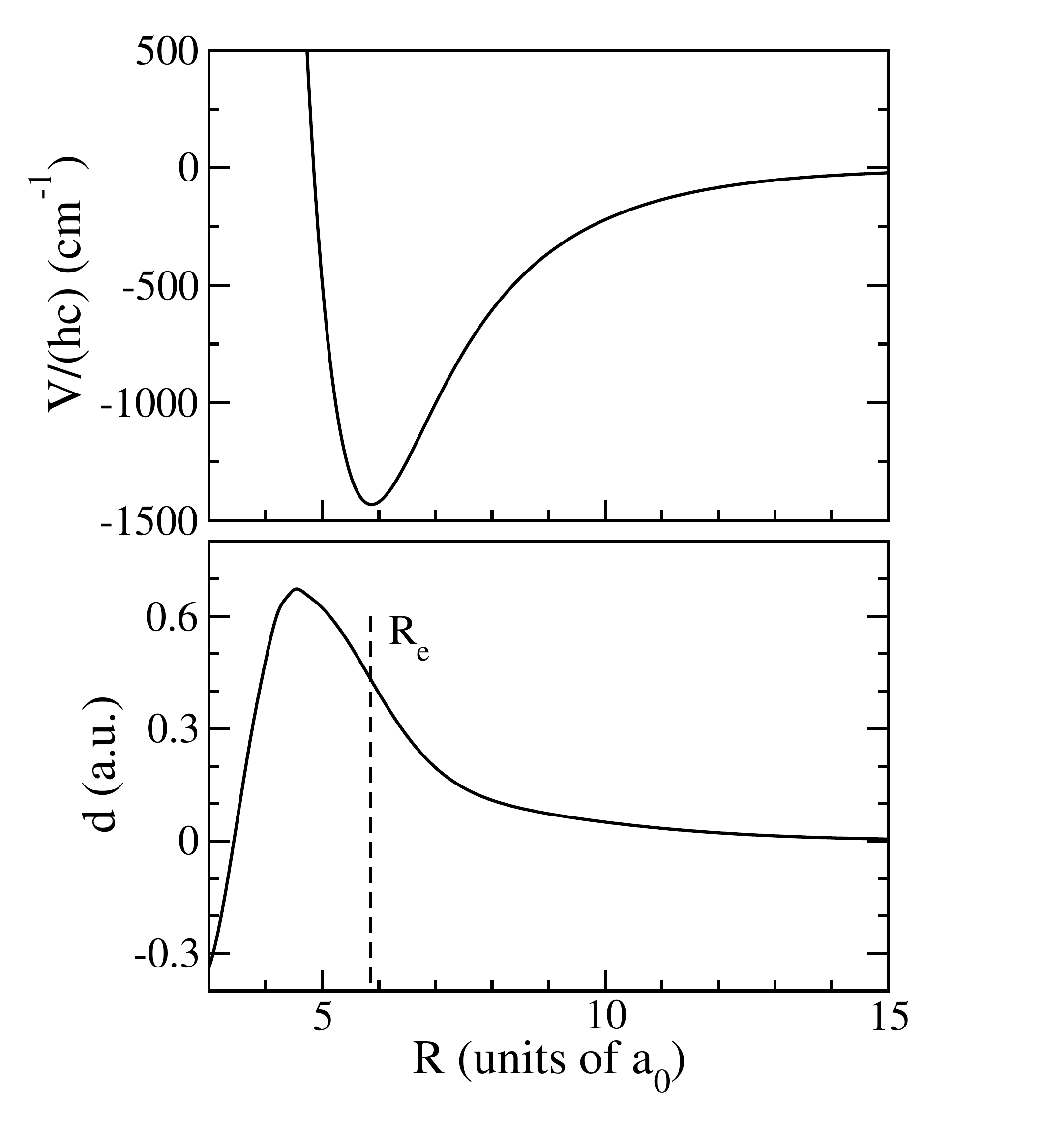}
\caption{Top panel: electronic X$^2\Sigma^+$ ground state potential of
LiMg; bottom panel: permanent dipole moment, where a vertical dashed line
indicates the equilibrium separation of the ground state potential; There
the dipole moment is 0.44 a.u., while the vibrationally averaged value
for the $v=0$ level is 0.413 a.u.. }
\label{limg}
\end{figure}

Here we apply RCCSD(T) and UCCSD(T) methods with two different basis
sets to calculate the potential and dipole moment of the LiMg ground state.
Augmented polarized core-valence (aug-cc-pCVQZ) basis sets with
Li:(15s9p5d3f1g)/[8s7p5d3f1g] and Mg:(19s15p6d4f2g)/[9s8p6d4f2g] functions
\cite{Dunning:89,Prascher} are used to calculate the ground state
potential energy at the RCCSD(T) level of theory, where all electrons in
both Li and Mg are correlated.  Aug-cc-pv5z-DK  basis sets  with Li:
(14s8p4d3f2g1h)/[6s5p4d3f2g1h] and Mg: (20s14p4d3f2g1h)/[7s6p4d3f2g1h]
\cite{Prascher} are applied in the UCCSD(T) calculation. The calculation is optimized
for scalar relativistic calculations  of the all-electron Douglas-Kroll
Hamiltonian. 
A comparison
with the molecular constants measured in Ref.~\cite{Berry} shows that
our $D_0$ and $R_e$ values  are  in good agreement with this experiment.
Figure \ref{limg} shows the $R$-dependent potential and dipole moment.

In Table~\ref{limgbound} we present a complete list of $N=0$ bound states
of the X$^2\Sigma^+$ ground state potential of  $^7$Li$^{24}$Mg obtained by a DVR
calculation. The zero-point corrected energy is found to
be $D_0/(hc)$=1330.05 cm$^{-1}$ in agreement with experimental
measurement \cite{Berry}.

\begin{table}[b]
\caption{All vibrational energies of the $^7$Li$^{24}$Mg X$^2\Sigma^+$ ground state potential 
for the $N=0$ rotational state.  The potential is based on the RCCSD(T) 
calculation.} 
\begin{ruledtabular}
\begin{tabular}{rcrc}
 $\nu$&Energy/($hc$) (cm$^{-1}$)& $\nu$&Energy/($hc$)(cm$^{-1}$)\\
 0  &  -1330.05 &  10  &  -216.21\\
 1  &  -1164.35 & 11  &  -164.06 \\
 2  &  -1010.96 & 12  &  -120.16 \\
 3  &  -870.00  & 13  &  -83.91 \\
 4  &  -741.52  & 14  &  -54.87\\
 5  &  -625.41  & 15  &  -32.62\\
 6  &  -521.43  & 16  &  -16.78\\
 7  &  -429.16  & 17  &  -6.89\\ 
 8  &  -348.00  & 18  &  -1.75\\
 9  &  -277.27  & 19  &  -0.12\\  
\end{tabular}
\end{ruledtabular}
\label{limgbound}
\end{table} 
 
\subsection{LiCa molecule}

Currently, there is no reliable experimental data for the ground state
potential of the LiCa molecule in the literature. However, preliminary
experimental results on the electronic properties of the ground-state
LiCa potential based on high resolution spectroscopy were presented
at the EGAS conference in 2010 \cite{Pashov}. They gave the molecular
constants $D_e/(hc)$ = 2607.8(100) cm$^{-1}$, $R_e = 6.3415(5) a_0$, and
$\omega_e/(hc)$ = 195.2 cm$^{-1}$  for $^7$Li$^{40}$Ca.  Earlier theoretical
predictions for the LiCa ground state give a dissociation energy of $D_e/(hc)$
= 2177.7 cm$^{-1}$ and $R_e = 6.65 a_0$ using density functional theory
\cite{Jones}. Later a configuration-interaction calculation involving
an effective pseudo-potential \cite{Allouche} predicted  $D_e/(hc)$ = 2355
cm$^{-1}$ and $R_e = 6.23 a_0$.  An all-electron $\it ab~initio$ study
of LiCa in Ref.~\cite{Russon} reported $D_e/(hc)$ = 1935.7 cm$^{-1}$ and $R_e
= 6.44 a_0$. Experimental two-photon ionization spectroscopy reported
in the same work found that $R_e$ equals to $6.3865a_0$. The substantial
discrepancy in the reported characteristics and lack of the experimental
data on the electronic structure of this molecule has initiated our study.

An accurate treatment of the LiCa molecule with presumably weak bonding
requires a massive inclusion of correlation effects. To satisfy this
requirement we apply two coupled-cluster approaches, UCCSD(T) and
UCCSDT, that allow us to introduce higher-order correlation effects.
To probe the influence of basis sets on the LiCa characteristics we
used two different basis sets for each atom. 
These are all-electron basis sets with quadruple-zeta quality.
Basis sets (def2-QZVPP) for Li:(15s6p2d1f)/[6s4p2d1f] and 
Ca:(24s18p6d3f)/[11s6p4d3f] functions are taken from Ref.~\cite{Weigend}. 
Basis sets (aug-cc-pCVQZ) with Li:(15s9p5d3f1g)/[8s7p5d3f1g]
\cite{Dunning:89} and Ca:(25s19p10d4f2g)/[10s9p7d4f2g] \cite{Koput}
functions include larger correlation expansions.

\begin{figure}[h]
\includegraphics[scale=0.4]{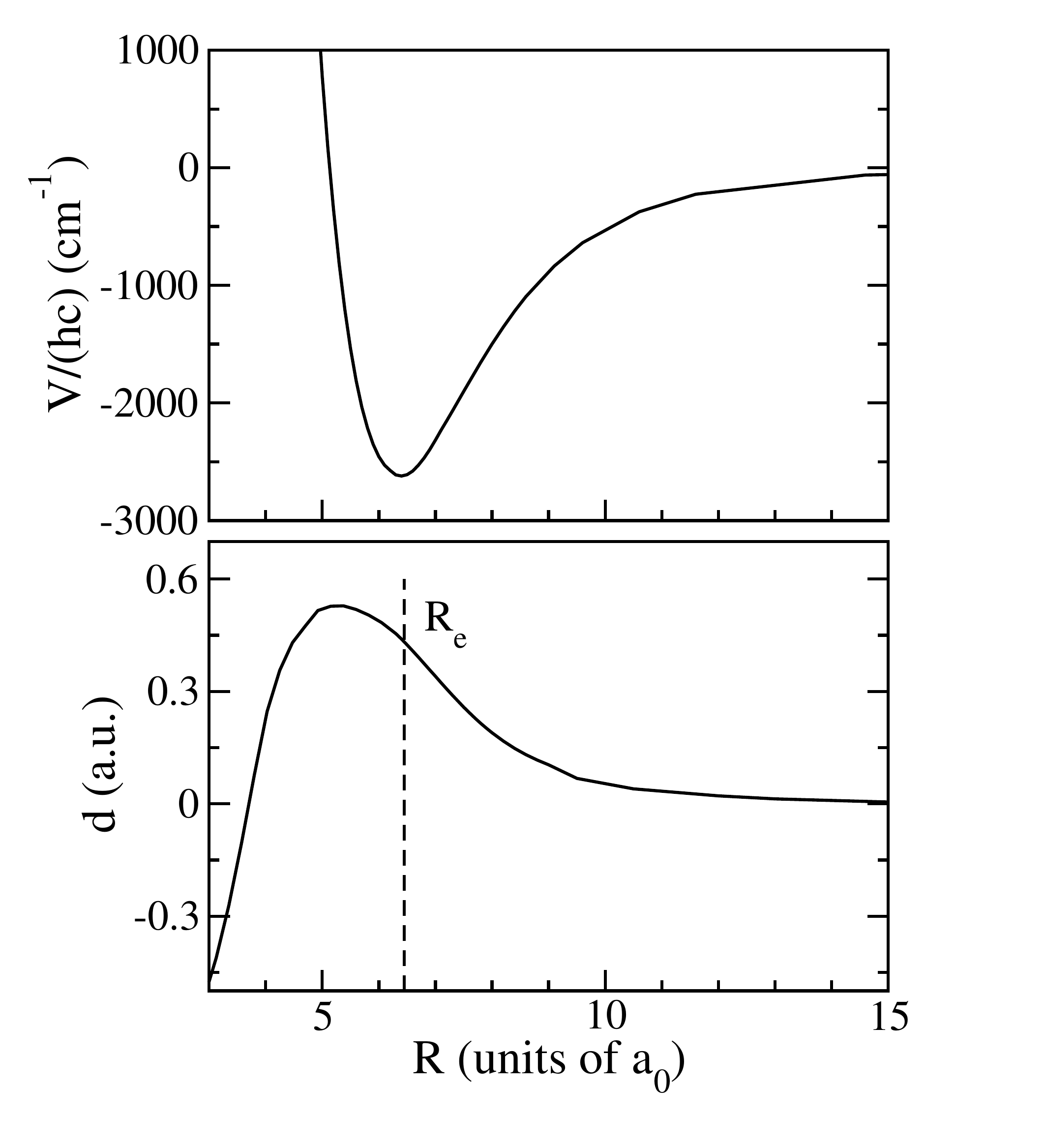}
\caption{Top panel: electronic ground state $^2\Sigma_{1/2}^+$ potential
of LiCa; bottom panel: permanent dipole moment of the ground state of
LiCa. The vertical dashed line indicates the equilibrium separation of the
ground state potential. At this separation the dipole moment is 0.044 a.u. 
The vibrationally averaged dipole moment with
the $v=0$ wave function is 0.437 a.u.}
\label{li_ca}
\end{figure}

We find that the molecular characteristics depend not only basis set but
also on the number of correlated electrons $n_{e\ell}$.  For example,
for a 3-electron correlation one $2s^1$ valence electron of Li and two
$4s^2$ valence electrons of Ca are correlated, whereas for 13-electron
correlation all electrons of Li and ten $3s^23p^64s^2$ electrons of
Ca are included in the active space.  
We conclude that using the UCCSD(T)
method and the more advanced basis set (aug-cc-pCVQZ) leads to an increase
of $D_e/(hc)$ and $d_e$ by $\approx$ 5\%.  For the UCCSDT method we expect
similar trends.  We believe that the UCCSDT calculation with the basis
set (aug-cc-pCVQZ) provides the most accurate electronic potential and
dipole moment. These are shown in Fig.~\ref{li_ca}.  Moreover, this
calculation is in a good agreement with the experimental molecular
constants reported by Ref.~\cite{Pashov}.

The upper panel of Fig.~\ref{li_ca} shows the potential energy curve of the X
$^2\Sigma^+$ state of LiCa as a function of interatomic separation. The
permanent dipole moment of this state as a function of $R$ is presented in the lower
panel.  A complete list of the $N=0$ vibrational 
energies of the ground state is presented in Table~\ref{licabound}.

\begin{table}[h]
\caption{All vibrational energies of the $^7$Li$^{40}$Ca X$^2\Sigma^+$ ground state potential 
for the rotational state $N=0$.  The potential is based on the UCCSD(T)/aug-cc-PCVQZ 
calculation.} 
\begin{ruledtabular}
\begin{tabular}{rcrc}
$\nu$&Energy/($hc$) (cm$^{-1}$)& $\nu$&Energy/($hc$) (cm$^{-1}$)\\
0  &    -2428.02 & 14  &  -442.18\\
1  &    -2232.09 & 15  &  -366.32\\
2  &    -2042.75 & 16  &  -298.77\\
3  &    -1861.87 & 17  &  -239.22\\
4  &    -1688.77 & 18  &  -187.31\\
5  &    -1524.31 & 19  &  -142.70\\
6  &    -1368.30 & 20  &  -105.03\\
7  &    -1221.05 & 21  &   -73.95\\
8  &    -1082.74 & 22  &   -49.08\\
9  &     -953.39 & 23  &   -30.02\\
10 &     -833.10 & 24  &   -16.33\\
11 &     -721.92 & 25  &    -7.41\\
12 &     -619.79 & 26  &    -2.47\\
13 &     -526.60 & 27  &    -0.42\\
\end{tabular}
\end{ruledtabular}
\label{licabound}
\end{table} 

\subsection{LiSr molecule}

To our knowledge there exist no data on the LiSr ground-state potential
based on experimental observation.  We are only able to compare our results
to the theoretical results of Ref.~\cite{Dulieu2010}. The calculation of
Ref.~\cite{Dulieu2010}  uses an effective core and core polarization
potentials that are incorporated in a configuration-interaction approach.
Their molecular constants are 
$R_e = 6.57 a_0$, $D_e/(hc)$ = 2587 cm$^{-1}$, $\omega_e/(hc)$ = 184.9 cm$^{-1}$,
$B_e/(hc)$ = 0.21 cm$^{-1}$, and $d_e$ = 0.13 a.u. In the present study 
we use different computational methods to predict these characteristics.

We perform calculations for LiSr that are similar to those for LiCa.
We use UCCSD(T) and UCCSDT methods with three different basis sets.
The inner electrons of the closed $1s^2 2s^2
...3d^{10}$ shells of Sr are chemically inactive and excluded
from the correlation calculation. 
The $1s^22s^1$ electrons of Li and the ten $5s^24s^24p^6$ electrons
of Sr, are included in the all-order correlation calculation. 

In the first calculation we use the basis set
def2-QZVPP with Li:(15s6p2d1f)/[6s4p2d1f] and Sr:(8s8p5d3f)/[7s5p4d3f] 
functions \cite{Weigend} and the Stuttgart ECP28MDF
relativistic effective core potential (ECP) \cite{Stoll}.
In a second calculation we use the basis set ``\cite{Lim}'' with Li: (14s8p4d2f)/[7s6p4d2f]
and Sr: (14s11p5d4f1g)/[ 8s8p5d4f1g] functions.  The effective core
potential is from Ref.~\cite{Lim}. 
Finally, we performed a calculation with
a more-advanced all-electron basis set (aug-cc-pCV5Z) described by Li:
(18s12p7d5f3g1h)/[10s9p7d5f3g1h] and Sr: (23s19p12d4f2g)/[11s10p7d4f2g]
functions from \cite{Prascher}. The UCCSDT calculation is time consuming
and has been performed for basis set ``\cite{Lim}'' only.

\begin{figure}
\includegraphics[scale=0.4]{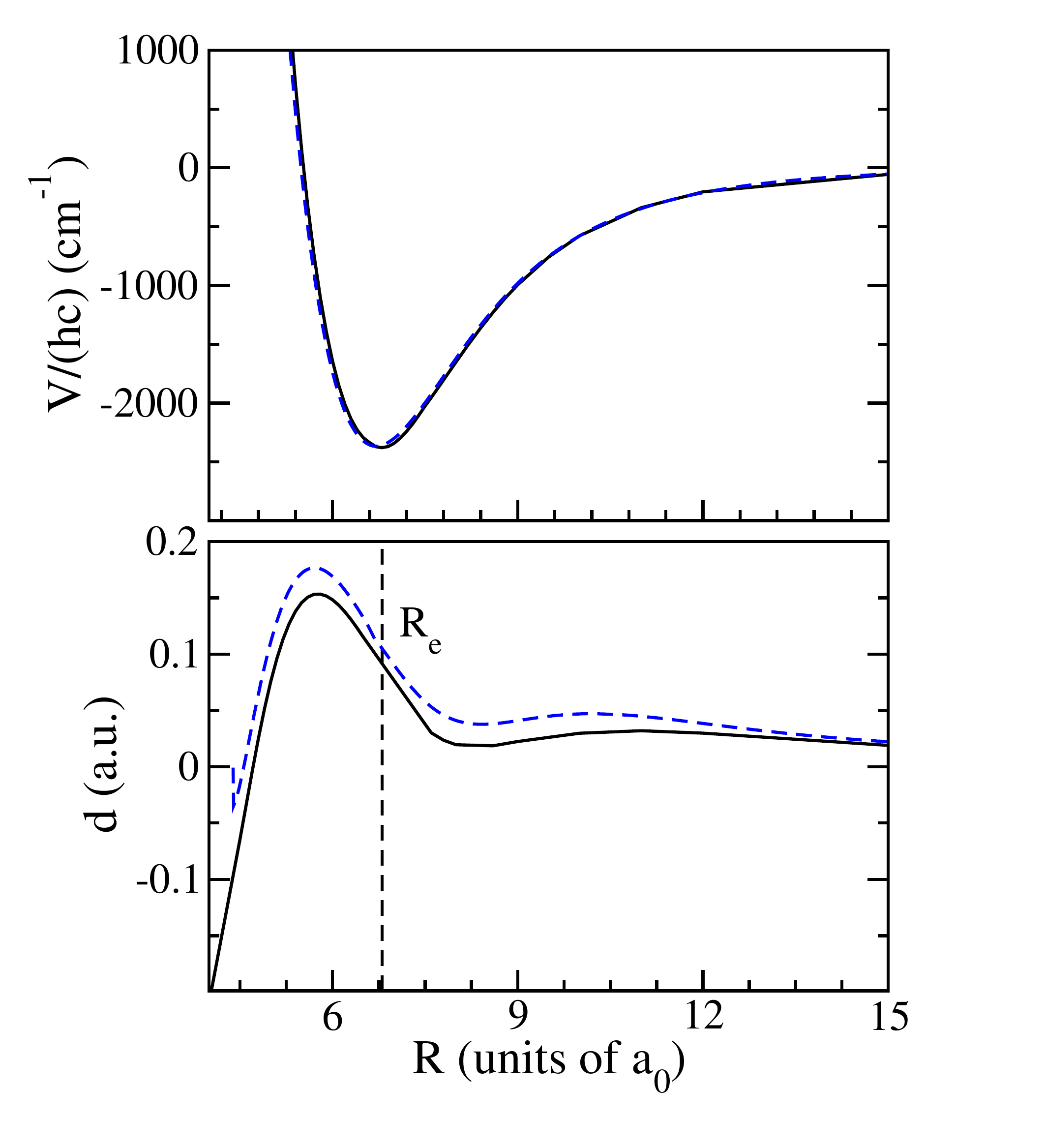}
\caption{Top panel: electronic ground-state $^2\Sigma+$ potential of LiSr;
bottom panel: permanent dipole moment of the ground state of LiSr.
The potentials and dipole moments are obtained by two different methods:
UCCSDT (solid curve) and  UCCSD(T) (dashed curve). The vertical dashed line 
in the bottom panel indicates the equilibrium separation of the ground state 
potential. The vibrationally averaged dipole moment with the $v=0$ 
wave function is 0.111 a.u.;
 }
\label{li_sr}
\end{figure}

Results of the
most advanced calculation by the UCCSDT method with the basis set ``\cite{Lim}'' and
the UCCSD(T) method with basis set (aug-cc-pCV5Z) are shown in Fig.~\ref{li_sr}.
We see that the potential energy of the two calculations nearly
coincide, whereas the dipole moment curves in the bottom panel are
slightly shifted relative to each other. We believe that this difference
is within the accuracy of our calculation.  The largest discrepancy of about
7\% with the results of Ref.~\cite{Dulieu2010} occurs at large internuclear separation,
whereas at $R_e$ the dipole moment $d_e$ differs by 2\% to 3\% only.

The DVR calculation of the bound state energy gives 
$D_0/(hc)$ = 2275.59 cm$^{-1}$ and $\omega/(hc)$ =182.2 cm$^{-1}$.
Table~\ref{lisrbound} lists all $N=0$ bound-state energies for $^7$Li$^{88}$Sr.

\begin{table}[h]
\caption{All vibrational energies of the $^7$Li$^{88}$Sr X$^2\Sigma^+$ ground state potential 
for the rotational state $N=0$.  The potential is based on the UCCSD(T) 
calculation with basis set \cite{Prascher}} 
\begin{ruledtabular}
\begin{tabular}{rcrc}
$\nu$&Energy/($hc$) (cm$^{-1}$)& $\nu$&Energy/($hc$) (cm$^{-1}$)\\
0   &   -2275.59  &15 &   -397.78\\
1   &   -2099.56  &16 &   -331.60\\
2   &   -1931.06  &17 &   -272.33\\
3   &   -1769.35  &18 &   -219.76\\
4   &   -1614.41  &19 &   -173.66\\
5   &   -1466.66  &20 &   -133.76\\
6   &   -1326.14  &21 &    -99.83\\
7   &   -1192.91  &22 &    -71.56\\
8   &   -1067.12  &23 &    -48.65\\
9   &    -948.89  &24 &    -30.80\\
10  &    -838.24  &25 &    -17.64\\
11  &    -735.17  &26 &     -8.73\\
12  &    -639.65  &27 &     -3.42\\
13  &    -551.65  &28 &     -0.86\\
14  &    -471.08  &29 &     -0.04\\
\end{tabular}
\end{ruledtabular}
\label{lisrbound}
\end{table}

\subsection{LiYb molecule}

Recently, two extensive quantum-chemical calculations of the ground and
excited states of the LiYb molecule have been reported by Zhang {\it et
al.} \cite{Zhang} and by Gopakumar {\it et al.} \cite{Gopakumar2010}.
The first calculation used both MRCI and UCCSD(T) methods. Scalar
relativistic effects were included by the Douglas-Kroll Hamiltonian
and an effective core potential. Their molecular constants for the
X$^2\Sigma^+$ ground state of $^7$Li$^{172}$Yb are the following: $R_e
= 6.681 a_0$, $D_e/(hc)$ = 1577 cm$^{-1}$, $\omega_e/(hc)$ = 147.36 cm$^{-1}$.
Different treatments of the Yb core electrons lead to different values
of the dipole moment.  A RECP-pseudo-potential calculation gave 
$d_e$ = 0.022 a.u., while an all-electron DKH3 calculation gave a
smaller $d_e$ of 0.011 a.u. Results of a calculation
by ~\cite{Gopakumar2010} were based on both the CASPT2 perturbation theory
and the CCSD(T) method. Relativistic effects were taken into account
through the Douglas-Kroll-Hess Hamiltonian. Their best values of molecular
constants are: $R_e = 6.669 a_0$, $D_e/(hc)$ = 1421.96 cm$^{-1}$, $\omega_e/(hc)$
= 135.54 cm$^{-1}$.

Our goal here is to perform an independent calculation of the ground state 
potential and permanent dipole moment of LiYb using the UCCSD(T) method with
another effective ECP28MWB pseudo-potential and with basis sets for Li:
(15s6p2d1f)/[6s4p2d1f] and Yb: (14s13p10d8f6g)/[10s8p5d4f3g] functions from
\cite{Cao}.

\begin{figure}
\includegraphics[scale=0.4]{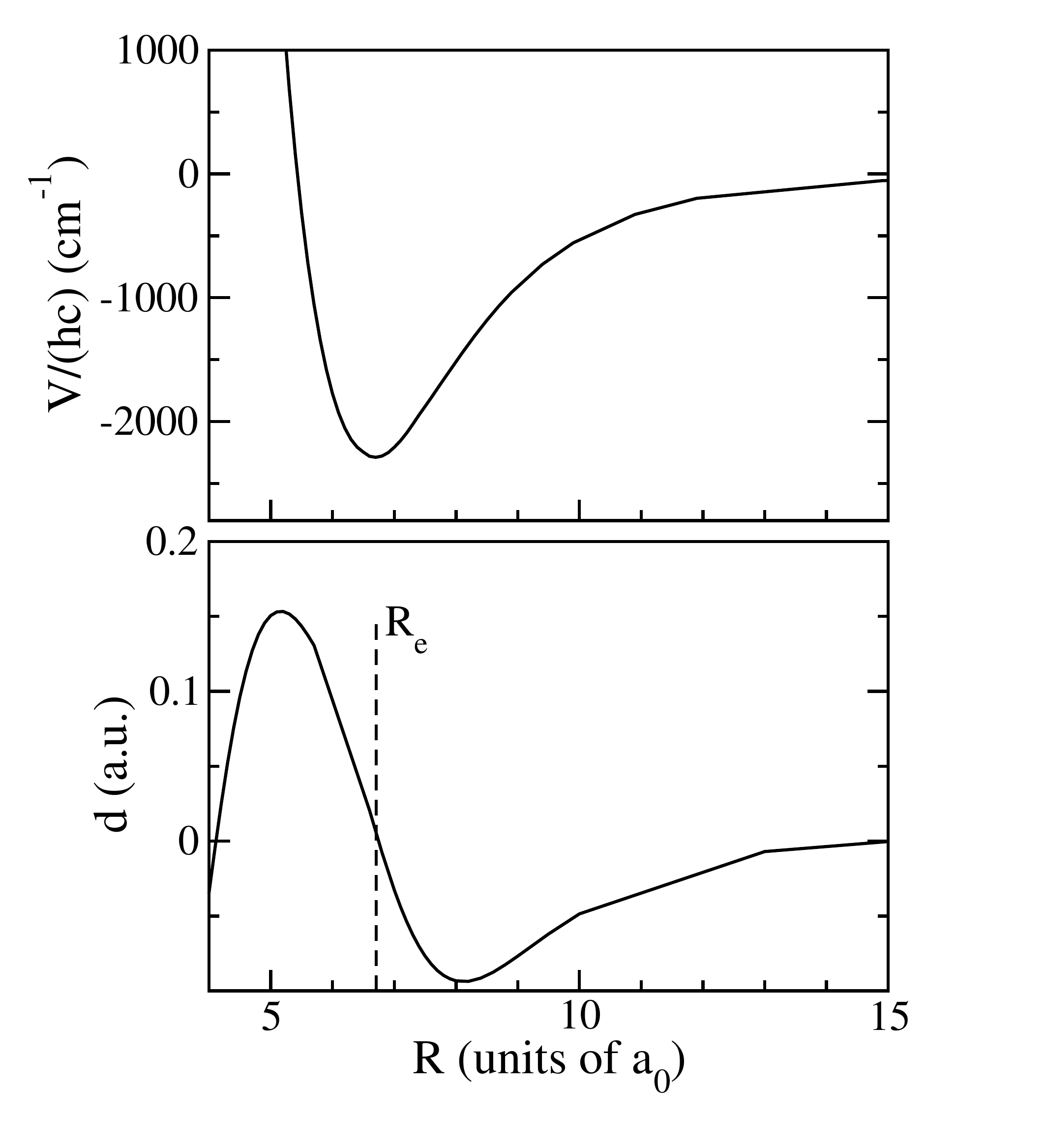}
\caption{Top panel: electronic ground-state $^2\Sigma^+$ potential of LiYb;
bottom panel: permanent dipole moment of the ground state.
The vertical dashed line in the bottom panel indicates
the equilibrium separation of the ground state potential;
}
\label{li_yb}
\end{figure}

Results of our calculation are shown in Fig.~\ref{li_yb}. The
upper panel of Fig.~\ref{li_yb} shows the ground state potential
of LiYb, which is significantly deeper than the potentials obtained in
Refs.~\cite{Zhang,Gopakumar2010}. The permanent dipole moment is
presented in the bottom panel. It is small and agrees well with the DKH3
value of Ref.~\cite{Zhang}. 

Table~\ref{liybbound} shows all $N=0$ bound-state energies for $^7$Li$^{172}$Yb.
\begin{table}[h]
\caption{All vibrational energies of the $^7$Li$^{172}$Yb X$^2\Sigma^+$ ground state potential
for the rotational state $N=0$.  The potential is based on the UCCSD(T)
calculation with basis set \cite{Cao}}
\begin{ruledtabular}
\begin{tabular}{rcrc}
$\nu$&Energy/($hc$) (cm$^{-1}$)& $\nu$&Energy/($hc$) (cm$^{-1}$)\\
0   &   -2277.12  &16 &   -349.63\\
1   &   -2103.98  &17 &   -289.41\\
2   &   -1938.15  &18 &   -235.73\\
3   &   -1778.86  &19 &   -188.38\\
4   &   -1626.11  &20 &   -147.10\\
5   &   -1480.30  &21 &   -111.67\\
6   &   -1341.45  &22 &    -81.82\\
7   &   -1209.63  &23 &    -57.26\\
8   &   -1084.99  &24 &    -37.70\\
9   &    -967.64  &25 &    -22.84\\
10  &    -857.62  &26 &    -12.36\\
11  &    -754.90  &27 &     -5.66\\
12  &    -659.49  &28 &     -1.94\\
13  &    -571.36  &29 &     -0.36\\
14  &    -490.44  &30 &     -0.005\\
15  &    -416.57  &   &          \\
\end{tabular}
\end{ruledtabular}
\label{liybbound}
\end{table}

\section{Summary}

The objective of our research has been a systematic investigation of the
X$^2\Sigma^+$ ground state properties of Li-group II molecules.
To achieve this goal we applied powerful quantum chemistry
UCCSD(T) and UCCSDT methods with large basis sets to calculate
potential energies and permanent dipole moments. Table~\ref{all_const}
displays molecular constants obtained in our calculations.

\begin{figure}
\includegraphics[scale=0.35]{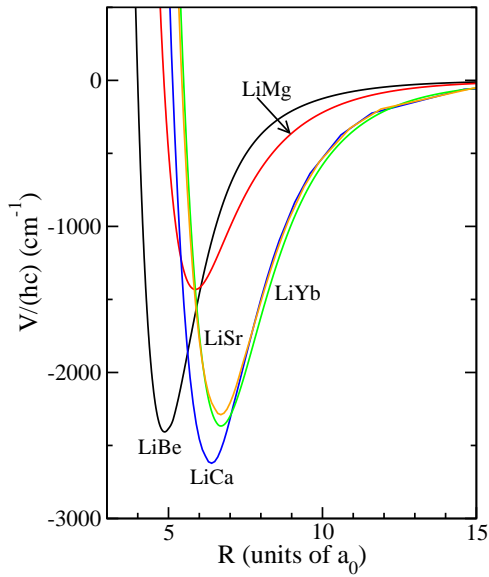}
\caption{Electronic ground state $^2\Sigma^+$ potentials of Li-group II molecules
and LiYb as a function of interatomic separation.}
\label{All_potentials}
\end{figure}

\begin{figure}
\includegraphics[scale=0.35]{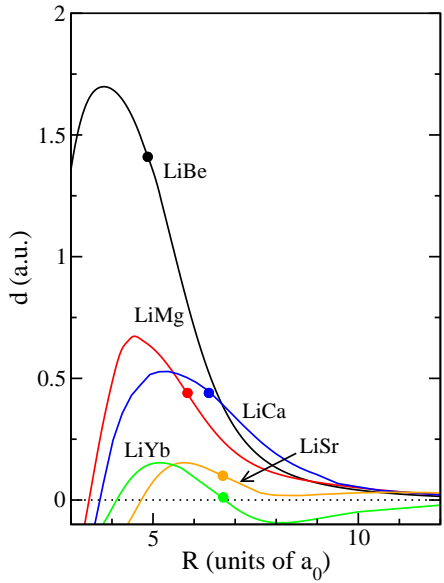}
\caption{Permanent dipole moment of the ground state of the Li-group II molecules
and LiYb as a function of interatomic separation. Filled circes indicate dipole
moment values at equiliblium separations of corresponding molecules.}
\label{All_dipoles}
\end{figure}

Figures~\ref{All_potentials} and \ref{All_dipoles} and Table~\ref{all_const}
summarize our results. Figure~\ref{All_potentials} reveals a most striking 
irregularity in the dissociation energy and $R_e$ among the Li-group II molecules.  
The deepest and most shallow potentials belong to the LiCa and LiMg molecules,
respectively.

Figure~\ref{All_dipoles} shows dipole moment functions for all molecules 
considered in our study. There is some consistency in the maximum value of $d$ with
the reduced mass of the molecule: heavier molecules have smaller dipole moment.  
The dipole moment averaged over the ground state vibrational wave function
is always a few tenths of atomic units smaller than $d_e$.  
The LiSr and LiYb  molecule have the smallest dipole moments
among all molecules in this study.

\begin{table}
\caption{Molecular constants for the X$^2\Sigma^+$ state of LiBe, LiMg, LiCa,
LiSr, and LiYb obtained by the RCCSD(T), UCCSD(T) and UCCSDT methods.
Here, 1 $a_0$=0.0529177 nm is the Bohr radius and a dipole moment of 1 a.u.
corresponds to 1 $ea_0$, where $e$ is the electron charge. Also $h$ is Planck's constant
and $c$ the speed of light.
} \label{all_const}
\vspace*{0.2cm}
\begin{ruledtabular}
\begin{tabular}{l|llll} 
Description   &$R_e$~~~~~ & $\omega_e/(hc)$~~~~~ & $D_e/(hc)$~~~~~  & $d_e$ \\
         & ($a_0$) & (cm$^{-1}$) & (cm$^{-1}$) & (a.u.) \\ \hline
\multicolumn{5}{c}{\bf $^7$Li$^9$Be } \\ \hline \hline
\multicolumn{1}{c|}{\bf UCCSD(T) }\\
 aug-cc-pV5Z-DK &4.873&299.5 & 2406& 1.41 \\
{\bf Experiment} \cite{Schlachta1990}
               & 4.894   &         &         &\\ \hline \hline
\multicolumn{5}{c}{ \bf $^7$Li$^{24}$Mg } \\ \hline
\multicolumn{1}{c|}{ \bf RCCSD(T)} \\
 aug-cc-pCVQZ   & 5.86  &  174.4   & 1417   &   0.32  \\ 
\multicolumn{1}{c|}{ \bf UCCSD(T)} \\
aug-cc-pV5Z-DK  &  5.87     &  206.1   & 1432    &  0.44  \\
{\bf Experiment}\cite{Berry}
                &   5.9     & 190      & 1423\footnote{This value is estimated from $D_0$ of Ref.~\cite{Berry}}
        &        \\ \hline \hline
\multicolumn{5}{c}{ \bf $^7$Li$^{40}$Ca } \\ \hline
\multicolumn{1}{c|}{ \bf UCCSD(T)} \\
def2-QZVPP   &  6.410   &  196.0    &  2320     &  0.438   \\
aug-cc-pCVQZ & 6.358  &  205.5    &  2460   &    0.456  \\ 
\multicolumn{1}{c|}{ \bf UCCSDT} \\
aug-cc-pCVQZ &  6.357     &  207.1    &   2607    &  0.440  \\
{\bf  Experiment }\cite{Pashov}
             & 6.3415(5)     & 195.2     &   2607.8(100)&   \\ \hline \hline
\multicolumn{5}{c}{ \bf $^7$Li$^{88}$Sr } \\ \hline
\multicolumn{1}{c|}{ \bf UCCSD(T)} \\
def2-QZVPP &6.766&167.0 &2165   & 0.109 \\
basis of \cite{Lim} &6.712 &183.0 &2302 & 0.117 \\
aug-cc-pCV5Z &6.700 &182.2&2367 & 0.112  \\
\multicolumn{1}{c|}{ \bf UCCSDT} \\
basis of \cite{Lim}  &6.711 & 184.2 & 2401 & 0.096 \\ \hline \hline
\multicolumn{5}{c}{ \bf $^7$Li$^{172}$Yb } \\ \hline
\multicolumn{1}{c|}{ \bf UCCSD(T)} \\
basis of \cite{Cao}  &6.710&181.5 &2289 & 0.011 
\end{tabular}
\end{ruledtabular}
\end{table}

Table~\ref{all_const} presents molecular constants calculated by the
RCCSD(T), UCCSD(T) and UCCSDT methods using several large basis sets
available in the literature.  Comparison of our most accurate (either UCCSD(T) or UCCSDT)
calculations with existing experimental measurements of the molecular
constants shows that we are able to reproduce the bond length ($R_e$) ,
harmonic frequency ($\omega_e$), and dissociation energy ($D_e$) within
5\%, 8\%, and 6\%, respectively.

The dipole moment as a function $R$ is calculated in terms of a charge transfer and induced
charge transfer. These two contributions have opposite signs over a large
range of internuclear separations. We conclude from our calculations
that charge transfer from the Li-atom to the alkaline-earth atom prevails at
all separations. This trend is reversed for LiYb at $R > 7 a_0$.

We also highlight the importance of the correlation effects in
calculations of the alkali-alkaline-earth molecules. We have gathered
numerical evidence to show that molecular constants of these molecules,
such as dissociation energy and permanent dipole moment of the ground
state, are extremely sensitive to high order correlations.  We believe that
the best values of these characteristics are obtained by using the UCCSDT
method, where correlation effects are taken into account at all orders.

Finally, we want to emphasize that LiBe and LiMg are not likely candidates
for ultracold studies because of the difficulty of cooling Be or Mg.
However, LiCa looks quite promising, since a Ca BEC \cite{Kraft2009}
has been made, and its dipole moment is comparable to the known RbCs
case \cite{Dulieu2005}.  The $^2\Sigma^+$ ground state gives an extra
spin degree of freedom relative to the $^1\Sigma^+$ RbCs species,
thus more quantum control possibilities.  While ultracold Li + Sr and
Li + Yb mixtures are possible for making the molecules, the dipole
moments of the ground state molecules are rather modest.  It remains
to be determined how best to make the molecules from the trapped atoms.
One possibility is to make ``preformed pairs'' in a dual species optical
lattice \cite{Damski2003, Freericks2010}, and then use intercombination
line photoassociation followed by STIRAP \cite{Ni2008}.  Although all
of these molecular species are expected to be highly chemical reactive
in their ground states since the formation of the dimers is exotermic
LiCa might have a large enough dipole moment to allow it to be protected
against destructive collisions by aligning it with an electric field in
a 1D optical lattice \cite{Zoller2006,JunYe2011,Julienne2011}.

\section{Acknowledgments}
The Temple University team and PSJ acknowledge support by an AFOSR MURI
grant on ultracold polar molecules. Work at Temple University was also
supported by  NSF Grant PHY-1005453, and J. K. would like to thank for 
the  financial support through NSF Grant CHE-0848110 to Millard H. Alexander.


\begin{references}
\bibitem{Zoller2006}A. Micheli, G. K. Brennen, and P. Zoller, Nature Phys. 2, 341 (2006).
\bibitem{sorensen}L. K. S$\o$rensen, S. Knecht, T. Fleig, and C. M. Marian, 
J. Phys. Chem. {\bf 113}, 12607 (2009).
\bibitem{Carr2009}L. D. Carr, D. DeMille, R. Krems, and J. Ye, New J. Phys. {\bf 11}, 055049 (2009).
\bibitem{Krems2010}J. Perez-Rios, F. Herrera, and R. V. Krems, New J. Phys. {\bf 12}, 103007 (2010). 
\bibitem{Hutson}P. S. {\.Z}uchowski, J. Aldegunde, and J. M. Hutson, Phys. Rev. Lett. {\bf 105},
153201 (2010). 
\bibitem{Dulieu2010}R. Gu{\'e}rout, M. Aymar, and O. Dulieu, Phys. Rev. A {\bf 82}, 042508 (2010).
\bibitem{Okano2010}M. Okano, H. Hara, M. Muramatsu, K. Doi, S. Uetake, Y. Takasu, and
Y. Takahashi, Appl. Phys. B: Lasers Opt. {\bf 98}, 691 (2010).
\bibitem{Zhang}P. Zhang, H. R. Sadeghpour, A. Dalgarno, J. Chem. Phys. {\bf 133}, 044306 (2010).
\bibitem{Gopakumar2010}G. Gopakumar, M. Abe, B. P. Das, M. Hada, and K. Hirao, J. Chem. Phys. 
{\bf 133}, 124317 (2010).
\bibitem{Gupta2011}V. V. Ivanov et al., Phys. Rev. Lett. {\bf 106}, 153201 (2011).
\bibitem{Pashov}M. Ivanova, A. Pashov, A. Stein,  H. Kn\"ockel, and E. Tiemann, 
(private communication) (2011).
\bibitem{Krems2006}T.V. Tscherbul and R.V. Krems, Phys. Rev. Lett. {\bf 97},
083201 (2006).
\bibitem{Hutson2002}J. J. Hudson, B. E. Sauer, M. R. Tarbutt, and E. A. Hinds, Phys. Rev. Lett. 
{\bf 89}, 023003 (2002).
\bibitem{Fedichev1996} P. O. Fedichev, Yu. Kagan, G. V. Shlyapnikov, and J. T. M.
Walraven, Phys. Rev. Lett. {\bf 77}, 2913 (1996).
\bibitem{Bohn1996}J. L. Bohn and P. S. Julienne, Phys. Rev. A {\bf 54}, R4637 (1996).
\bibitem{Fatemi2000} F. K. Fatemi, K. M. Jones, and P. D. Lett, Phys. Rev. Lett.
{\bf 85}, 4462 (2000).
\bibitem{Theis2004}M. Theis, G. Thalhammer, K. Winkler, M. Hellwig, G. Ruff, R. Grimm, and
J. H. Denschlag, Phys. Rev. Lett. {\bf 93}, 123001 (2004).
\bibitem{Ciurylo2005}R. Ciurylo, E. Tiesinga, and P. S. Julienne, Phys. Rev. A {\bf 71}, 030701
 (2005).
\bibitem{Thalhammer2005}G. Thalhammer, M. Theis, K. Winkler, R. Grimm, J. H. Denschlag, 
Phys. Rev. A {\bf 71}, 033404 (2005).
\bibitem{JonesReview2006}K. M. Jones, E. Tiesinga, P. D. Lett, and P. S. Julienne, Rev. Mod. Phys.
{\bf 78}, 483 (2006).
\bibitem{Enomoto2008}K. Enomoto, K. Kasa, M. Kitagawa, and Y. Takahashi, Phys. Rev. Lett.
{\bf 101}, 203201 (2008).  
\bibitem{Pichler}G. Pichler, A. M. Lyyra, P. D. Kleiber, W. C. Stwalley, R. Hammer, 
and K. M. Sando, Chem. Phys. Lett. {\bf 156}, 467 (1989).
\bibitem{Berry}K. R. Berry and M. A. Duncan, Chem. Phys. Lett. {\bf 279}, 44 (1997).
\bibitem{Haeffler}G. Haeffler, D. Hanstorp, I. Kiyan, A. E. Klinkmueller, U. Ljungblad, 
D. J. Pegg, Phys. Rev. A {\bf 53}, 4127 (1996).
\bibitem{Bauschlicher}C. W. Bauschlicher, Jr. S. R. Langhoff, and H. Partridge, J. Chem. Ph
ys. {\bf 96}, 1240 (1992).
\bibitem{CFOUR}J.F. Stanton {\it et al.}, J. Chem. Theor. Comp. 4, 64 (2008).
\bibitem{MOLPRO} H.-J. Werner {\it et al.}, MOLPRO, Version 2008.1, a package of ab initio programs.
\bibitem{Lee1989}T. J. Lee {\it et al.} Theor. Chem. Acta {\bf 75}, 81 (1989).
\bibitem{Jones}R. O. Jones  J. Chem. Phys. {\bf 72}, 3197 (1980).
\bibitem{Fischer1991}I. Fischer, V. E. Bondybey, P. Rosmus, and H.-J. Wener, Chem Phys. {\bf 151}, 295 (1991).
\bibitem{Marino1992}M. M. Marino and W. C. Ermler, J. Chem. Phys. {\bf 96}, 3756 (1992).
\bibitem{Schlachta1990}R. Schlachta, I. Fischer, P. Rosmus, and V. E. Bondybey, Chem. Phys. Lett. 
{\bf 170}, 485 (1990).
\bibitem{Boys:70} S. F. Boys and F. Bernardi, Mol. Phys. {\bf 19}, 553 (1970).
\bibitem{Dunning:89} T.H. Dunning, Jr. J. Chem. Phys. {\bf 90}, 1007 (1989). 
\bibitem{Prascher}B. Prascher, D.E. Woon, K.A. Peterson, T.H. Dunning, Jr., and A.K. Wilson, 
Theor. Chem. Acc.,{\bf 128},69-82 (2011)
\bibitem{Colbert}D.T.Colbert and W.H.Miller, J. Chem. Phys. {\bf 96}, 1982 (1992).
\bibitem{Koput}J. Koput and K.A. Peterson, J. Phys. Chem. A {\bf 106}, 9595 (2002).
\bibitem{Allouche}A. R. Allouche and M. Aubert-Frecon, Chem. Phys. Lett. {\bf 222}, 524 (1994).
\bibitem{Russon}L. M. Russon, G. K. Rothschopf, M. D. Morse, A. I. Boldyrev, and J. Simons,
J. Chem. Phys. {\bf 109}, 6655 (1998).
\bibitem{Stoll}I. Lim, P. Schwerdtfeger, B. Metz, and H. Stoll, J. Chem. Phys. {\bf 122},
104103 (2005).
\bibitem{Weigend}F. Weigend and R. Ahlrichs, Phys.  Chem. Chem. Phys. {\bf 7}, 3297 (2005).
\bibitem{Lim}I.S. Lim, H. Stoll, P. Schwerdtfeger, J. Chem. Phys. {\bf 124}, 034107 (2006).
\bibitem{Dolg}M. Dolg, H. Stoll, H. Preuss, J. Chem. Phys. {\bf 90}, 1730 (1989).
\bibitem{Cao}X. Cao, M. Dolg, J. Molec. Struct. Theochem. {\bf 581} 139, (2002).
\bibitem{Kraft2009}S. Kraft {\it et al.}, Phys. Rev. Lett. {\bf 103}, 130401 (2009).
\bibitem{Dulieu2005}M. Aymar and O. Dulieu, J. Chem. Phys. {\bf 122}, 204302 (2005).
\bibitem{Damski2003}B. Damski, L. Santos, E. Tiemann, M. Lewenstein, S. Kotochigova, 
P. Julienne, and P. Zoller, Phys. Rev. Lett. {\bf 90}, 110401 (2003).
\bibitem{Freericks2010}J. K. Freericks, M. M. Maśka, Anzi Hu, Thomas M. Hanna, C. J. Williams, and P. S. Julienne, and R. Lemański, Phys. Rev. A {\bf 81}, 011605(R) (2010).
\bibitem{Ni2008}K.-K. Ni, S. Ospelkaus, M. H. G. de Miranda, A. Pe’er, B. Neyenhuis, 
J. J. Zirbel, S. Kotochigova, P. S. Julienne, D. S. Jin, J. Ye, Science {\bf 322}, 231
(2008).
\bibitem{JunYe2011}M. H. G. de Miranda, A. Chotia, B. Neyenhuis, D.Wang, 
G. Quéméner, S. Ospelkaus, J. L. Bohn, J. Ye, and D. S. Jin, Nature Phys. 
{\bf 7}, 502 (2011).
\bibitem{Julienne2011}P. S. Julienne, T. M. Hanna, and Z. Idziaszek, Phys. Chem. Chem.
Phys. Advanced Article DOI:10.1039/C1CP21270B.
\end{references}
\end{document}